\documentclass[sigconf]{acmart}
\AtBeginDocument{%
  }
\copyrightyear{2025}
\acmYear{2025}
\setcopyright{rightsretained}
\acmConference[CHI EA '25]{Extended Abstracts of the CHI Conference on
Human Factors in Computing Systems}{April 26-May 1, 2025}{Yokohama, Japan}
\acmBooktitle{Extended Abstracts of the CHI Conference on Human Factors in Computing Systems (CHI EA '25), April 26-May 1, 2025, Yokohama,
Japan}\acmDOI{10.1145/3706599.3720269}
\acmISBN{979-8-4007-1395-8/2025/04}

\usepackage{hyperref}
\usepackage{enumitem}
\usepackage{subcaption}
\usepackage{makecell}
\usepackage{longtable}
\usepackage{array}
\usepackage{graphicx}

\begin{document}

\title[Redefining Research Crowdsourcing]{Redefining Research Crowdsourcing: Incorporating Human Feedback with LLM-Powered Digital Twins
}

\author{Amanda Chan}
\authornote{All four authors contributed equally to this research.}
\affiliation{%
  \institution{Princeton University}
  \city{Princeton}
  \state{New Jersey}
  \country{USA}
  \postcode{08540}
}
\email{ac2921@princeton.edu}

\author{Catherine Di}
\authornotemark[1]
\affiliation{%
  \institution{Princeton University}
  \city{Princeton}
  \state{New Jersey}
  \country{USA}
  \postcode{08540}
}
\email{cathydi@princeton.edu}

\author{Joseph Rupertus}
\authornotemark[1]
\affiliation{%
  \institution{Princeton University}
  \city{Princeton}
  \state{New Jersey}
  \country{USA}
  \postcode{08540}
}
\email{joerup@princeton.edu}

\author{Gary Smith}
\authornotemark[1]
\affiliation{%
  \institution{Princeton University}
  \city{Princeton}  
  \state{New Jersey}
  \country{USA}
  \postcode{08540}
}
\email{garysmith@princeton.edu}

\author{Varun Nagaraj Rao}
\affiliation{%
  \institution{Princeton University}
  \city{Princeton}
  \state{New Jersey}
  \country{USA}
  \postcode{08540}
}
\email{varunrao@princeton.edu}

\author{Manoel Horta Ribeiro}
\affiliation{%
  \institution{Princeton University}
  \city{Princeton}
  \state{New Jersey}
  \country{USA}
  \postcode{08540}
}
\email{manoel@princeton.edu}

\author{Andrés Monroy-Hernández}
\affiliation{%
  \institution{Princeton University}
  \city{Princeton}
  \state{New Jersey}
  \country{USA}
  \postcode{08540}
}
\email{andresmh@princeton.edu}

\renewcommand{\shortauthors}{Chan, Di, Rupertus, and Smith et al.}

\begin{abstract}
  Crowd work platforms like Amazon Mechanical Turk and Prolific are vital for research, yet workers' growing use of generative AI tools poses challenges. Researchers face compromised data validity as AI responses replace authentic human behavior, while workers risk diminished roles as AI automates tasks. To address this, we propose a hybrid framework using digital twins, personalized AI models that emulate workers' behaviors and preferences while keeping humans in the loop. We evaluate our system with an experiment ($n$=88 crowd workers) and in-depth interviews with crowd workers ($n$=5) and social science researchers ($n$=4). Our results suggest that digital twins may enhance productivity and reduce decision fatigue while maintaining response quality. Both researchers and workers emphasized the importance of transparency, ethical data use, and worker agency. By automating repetitive tasks and preserving human engagement for nuanced ones, digital twins may help balance scalability with authenticity.
\end{abstract}

\ccsdesc[500]{Human-centered computing~Empirical studies in HCI}
\ccsdesc[500]{Information systems~Crowdsourcing}
\ccsdesc[500]{Information systems~Collaborative and social computing systems and tools}

\keywords{digital twin, crowd work, AI uncertainty, MTurk, Prolific}

\begin{teaserfigure}
  \includegraphics[width=\textwidth]{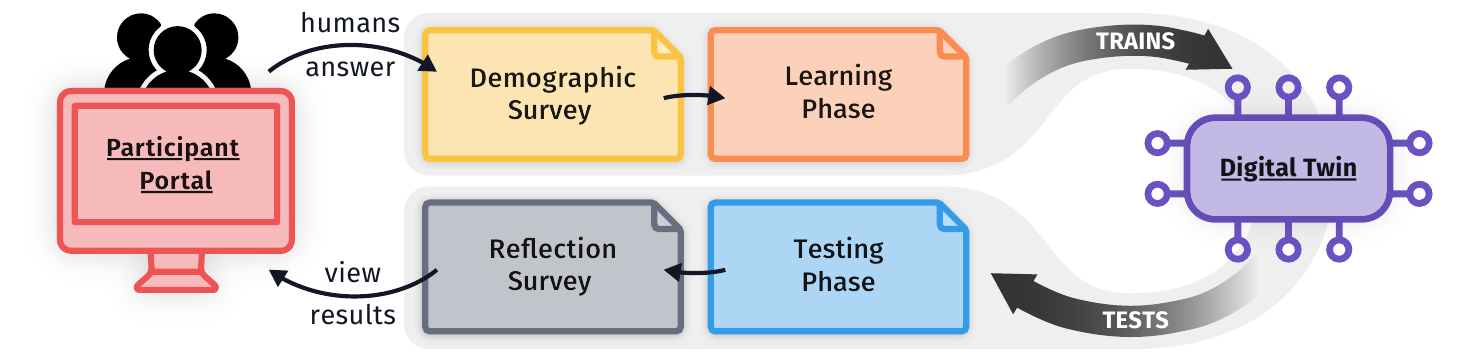}
  \Description[Overview of the participant study process with the Digital Twin system.]{
    Diagram showing the participant study process: participants (i.e., crowd workers) access the Participant Portal to answer the Demographic Survey, proceed to the Learning Phase, and move to the Testing Phase. The Digital Twin system is trained using the learning data and tested during the Testing Phase. After completing the study, participants answer the Reflection Survey to provide feedback on the process before the researcher can view the final results.}
  \caption{Participants complete surveys to train their digital twin and test its ability to answer questions in their place.}
  \label{fig:teaser}
\end{teaserfigure}

 \maketitle
\section{Introduction}
Crowd work on online labor market platforms, such as Amazon Mechanical Turk (MTurk) and Prolific, is widely used in both industry and academia, supporting machine learning dataset creation as well as social science research examining human behavior ~\cite{gray2019ghost,salganik2019bit}. This paper focuses on crowd work in social science research, where the growing accessibility of Large Language Models (LLMs), such as ChatGPT and Gemini, has significantly reshaped the ecosystem \cite{wu2023llms}. Crowd workers increasingly utilize LLMs, perhaps due to the precarious nature of crowd work  \cite{gray2019ghost}, often in ways that are difficult to detect~\cite{gilardi2023chatgpt,veselovsky_artificial_2023,veselovsky_prevalence_2023}.

These developments raise critical concerns for the researchers running the studies and the crowd workers completing them.
For researchers, the growing use of LLMs by crowd workers threatens the validity of their research, which may reflect \textit{AI} rather than \textit{human} behavior.
For crowd workers, LLM's growing capabilities threaten their livelihood by potentially reducing the demand for human labor in crowd work. Moreover, reliance on LLMs diminishes the value of their contributions as responses become less reliable and valuable for researchers.

Concerns about the quality of data generated by humans (often through crowd work) have led to a push for the creation of high-quality synthetic data ~\cite{businessinsider2024synthetic}.
Within this ``synthetic framework,'' machine learning models and social science studies alike would be powered by synthetic outputs created by LLMs~\cite{argyle_out_2022}.
Perhaps unsurprisingly, this approach has been met with widespread criticism, as synthetic data generated by humans fails to capture the nuances of human text and behavior~\cite{wang2024large,padmakumar2024doeswritinglanguagemodels}. 

To address these challenges, we propose a personalized, ``hybrid,'' human- and AI-assisted framework for crowd working that benefits researchers and crowd workers. This approach introduces LLM-powered ``digital twins'' that accurately emulate workers' behaviors and preferences, responding only when confident, to preserve the integrity of individual contributions while keeping humans in the loop.
We evaluated the system with 88 crowd workers, collecting feedback on their comfort and experience. Additionally, we interviewed five crowd workers and four social science researchers as key stakeholders to explore their experiences and perspectives on the system's potential impact.

Our study reveals key considerations for integrating AI into human-centered research while preserving research integrity and worker agency. Researchers were generally skeptical about AI's role in their studies, while crowd workers were less so. However, both agreed that the thoughtful implementation of digital twins could standardize response quality and reduce worker fatigue on repetitive tasks, making crowd work more accessible.

This work underscores the need for AI systems that augment, not replace, human judgment in research and data collection. It also highlights the importance of clear boundaries, transparency, and preserving human agency in AI-automated workflows. By prioritizing ethical and inclusive practices, our findings offer a framework for responsible AI integration that advances innovation while safeguarding human involvement.

\section{Related Work}

\subsection{Digital Twins and Personalized LLMs}
Digital twins (DTs) are virtual models of natural, engineered, or social systems that replicate their structure, context, and behavior while continuously updating with data from their physical counterparts~\cite{aiaa_digital_2020}. Although DTs have been applied in fields like civil engineering and medicine~\cite{babanagar_digital_2025, nagaraj_augmenting_2023}, their use in crowd working to model human behavior and decision-making is still emerging. By prompting LLMs to emulate human preferences, personalized human DTs, or generative agents, have shown significant promise in accurately replicating human attitudes and behaviors \cite{park_generative_2023, park_generative_2024}. Past work also indicates that DTs can improve decision-making by aligning AI systems with human preferences \cite{huang_simulation_2024}. In the context of crowd work, companies like Karya~\cite{karya} and Qloo~\cite{qloo} are exploring AI-driven market research. Still, it is unclear whether they capture individual nuances or rely on broader demographic trends \cite{kanda_efficient_2022, ibm_dt}. These findings underscore the potential of digital twins when tailored to individual opinions, emphasizing the need for \textit{personalized}, rather than generic, results. By focusing on individual-level LLM customization, this study seeks to improve the alignment and relevance of AI-assisted crowd work.

\subsection{Worker-AI Collaboration in Crowd work}
Integrating AI tools like LLMs into crowd work offers opportunities and challenges. While some researchers have demonstrated LLMs' ability to simulate human responses accurately \cite{argyle_out_2022, wu2023llms} and their potential for optimizing market research \cite{draghici_revolutionizing_2023}, others have cautioned that LLM-generated data may misrepresent social demographic groups (such as those defined by race, gender, or socioeconomic status), raising ethical concerns about authenticity and fairness \cite{wang_interpretable_2022}. Additionally, the widespread use of LLMs among crowd workers complicates the situation. Researchers have found that 33-46\% of workers used LLMs for text summarization tasks, and even when explicitly asked to avoid LLM assistance, nearly half continued to use them covertly \cite{veselovsky_artificial_2023, veselovsky_prevalence_2023}. This creates a dilemma: replacing humans with LLMs risks compromising diversity and authenticity, while traditional crowd work often involves generic, LLM-generated responses that fail to reflect individual human perspectives.
Prior work has explored human-AI collaboration strategies to address these challenges that include: real-time AI feedback systems to improve crowd worker performance \cite{wang_interpretable_2022}, dynamic task allocation to optimize task distribution between human and AI workers \cite{kobayashi_humanai_2021}, new methods to evaluate AI-worker collaboration  \cite{kanda_efficient_2022}, and human verification of LLM outputs to improve annotation accuracy \cite{wang_human-llm_2024}. Building on these promising results, we propose and evaluate a digital twin system to enhance crowd worker productivity while preserving human agency and authenticity.

\section{System Implementation}
\label{sec:sys-impl}
\noindent
We created a sample online research platform to simulate the use of digital twins to assist crowd workers in survey-answering tasks. The platform is a React application that interacts with OpenAI's GPT-4o model, an LLM, to create ``digital twin'' responses based on participant data.

\subsection{Demographic Data Gathering}
The system starts by asking crowd workers to answer 25 demographic questions (see Appendix \ref{survey-questions}), including age, gender, education, location, and political views, to contextualize the digital twin's responses based on their own characteristics. These responses serve as foundational data for tailoring the digital twin's behavior to the individual participant. 

\begin{figure}
    \centering
    \textbf{Learning Phase Interface}\\[0.5em]
    \includegraphics[width=\linewidth]{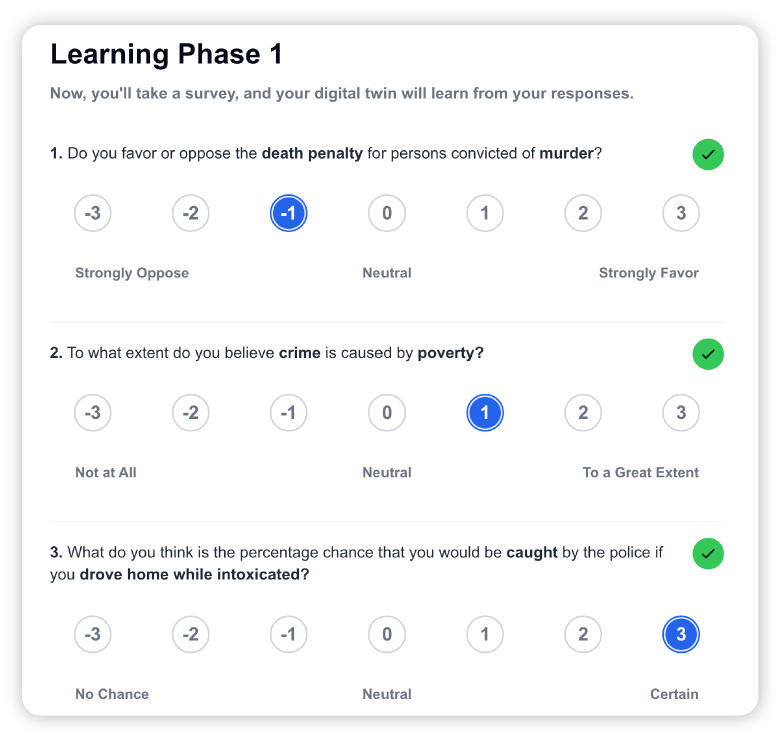}
    \Description[Crowd Worker Portal Screens]{The figure illustrates the Crowd Worker Portal interface used during the survey process. This Figure is for the Learning Phase 1 survey interface, featuring Likert-scale questions where participants rate their interest or importance on a scale ranging from -3 (i.e., strongly uninterested or very unimportant) to 3 (i.e., strongly interested or very important).}
    \caption{The learning survey interface design with Likert-scale questions.}
    \label{fig:example}
\end{figure}

\subsection{Learning Phase}
The crowd worker completes three learning surveys, each with 15 to 19 questions adapted from six social science surveys (see Appendix \ref{survey-questions}), evaluated on a 7-point Likert scale from -3 to +3. These surveys, selected from Hewitt and colleagues' compilation of 70 U.S. studies \cite{hewitt_predicting_2024}, assess attitudes, beliefs, and perceptions on social, political, and ethical issues. After submitting the responses, the system prompts the LLM with the same questions to predict the participant's answers based on demographic data and prior responses (see Appendix \ref{prompts}). The LLM's predictions are not shown to the user. Each phase iteratively refines the LLM's ability to simulate personalized answers.

\subsection{Testing Phase}
The system prompts the LLM with 43 additional survey questions from the same social science surveys as in the Learning Phase (see Appendix \ref{survey-questions}) and instructs it to predict the participant's answers based on their demographic survey responses and learning survey responses. The OpenAI API response includes a list of probabilities corresponding to each token in the LLM's output. For each question, the system finds the token corresponding to the numerical Likert-scale prediction, extracts the logprob metric associated with it, and calculates the linear probability. This acts as a measure of the LLM's confidence in its answer. For a given question, if the confidence is above a certain threshold\footnote{We empirically determined the 75\% threshold for accepting digital twin answers as roughly the optimal point for maintaining accuracy while enabling the digital twin to answer enough questions to be useful.} (75\%), the system accepts the answer and automatically fills out the survey question. The question will then appear to the participant as having been automatically completed by their ``digital twin.'' If the confidence is below the threshold, the system rejects the LLM answer and defers to the participant to answer manually.

\begin{figure}
    \centering
    \textbf{Testing Phase Interface}\\[0.5em]
    \includegraphics[width=\linewidth]{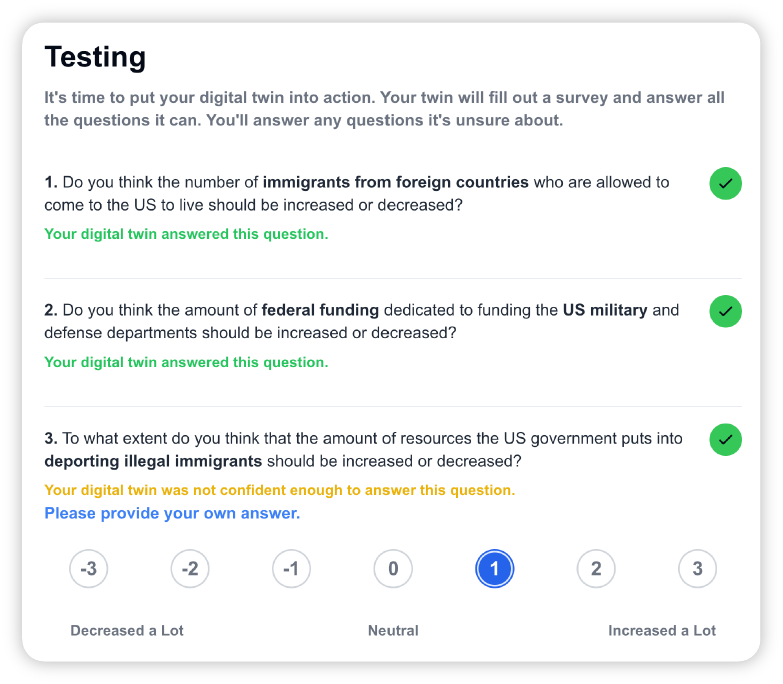}
    \Description[Participant Portal Screens]{The figure illustrates the Participant Portal interface used during the testing phase, where the Digital Twin attempts to answer survey questions based on prior learning. Participants provide their own answers for questions the Digital Twin is uncertain about, indicated with the prompt: Please provide your own answer. The interface includes clear feedback on which questions were answered by the Digital Twin and which require participant input.}
    \caption{The testing survey interface with both digital twin and human responses.}
    \label{fig:example}
\end{figure}

\subsection{Reflection Survey}
The participant independently answers all questions from the testing phase that were previously answered with high confidence ($>$75\%) by the LLM.%
\footnote{The reflection survey is intended to measure the LLM's accuracy in replicating individual participants' opinions, but it would not be included in a real system intended for use by crowd workers, since the intent is for the LLM to answer the questions automatically.} The LLM's prediction is only visible after the participant gives their own answer.
Finally, the participant answers 19 questions about the system (see Appendix \ref{survey-questions}), providing feedback on the digital twin's usability, accuracy, trustworthiness in decision-making, and overall user experience and satisfaction.

\section{Methods}

We recruited crowd workers ($n=88$) to test our system's accuracy and usability and to understand their reactions to it. Further, we interviewed both crowd workers across the U.S. ($n=5$) and social science researchers in the northeastern U.S. ($n=4$) to better understand their perspectives on using AI to augment crowd work. This study was approved by Princeton University's Institutional Review Board (IRB\# 17302), and all participants provided informed consent before participation.

\subsection{Platform Study with Crowd Workers} 
We initially recruited 105 participants on Prolific, who were directed to our online platform. After excluding incomplete responses, the final sample size consisted of 88 participants. These individuals spent approximately 20–25 minutes completing surveys on the system and were compensated \$5 for their time. Participants represented diverse backgrounds, all residing in the U.S. and originating from 32 different states. 54\% reported that they rely on crowd work as their main source of income, while 46\% have additional jobs. The majority of participants fell within the 26 to 45 age range. 58.1\% identified as White, 22.1\% as African American or Black, 11.6\% as Hispanic/Latinx, and 5.8\% as Asian. 54\% identified as male and 46\% identified as female.

The participants used the study platform (see Section \ref{sec:sys-impl}), completing learning and testing surveys to train and evaluate their digital twin. The survey questions addressed six key topics: immigration/race, crime, health, politics, terrorism, and ethics. These topics were carefully selected to provide a comprehensive evaluation of the digital twin system's ability to replicate human responses across diverse and nuanced subject areas. All survey questions were sourced from original studies\footnote{Some phrasing was slightly modified to adapt to a Likert scale format.} that had been previously administered to human crowd workers. The survey questions were distributed such that 55\% were part of the Learning Phase, while 45\% were included in the Testing Phase. By covering all six topics in both phases, the digital twin system progressively familiarized itself with participant responses, mimicking the iterative learning process of humans.

\subsection{Interviews with Crowd Workers and Researchers}
We interviewed five crowd workers (recruited via Reddit and Prolific) and four social science researchers (recruited through personal connections). The crowd workers were all active users of Prolific and/or Amazon Mechanical Turk, and the researchers all used crowd working platforms in their research. Interviews were conducted virtually, lasting about one hour each. Crowd workers received \$20 compensation, while researchers participated voluntarily. 

The interviews explored participants' experiences with crowd working platforms and AI's influence on task quality, alongside a walkthrough of the digital twin system. Feedback focused on design, ethical concerns, privacy, and the long-term implications of AI in crowd work. Discussions compared AI-only, hybrid, and human systems in terms of cost, workload, authenticity, and fairness.

We audio-recorded and transcribed all interviews, then conducted thematic analysis using open coding. We coded the transcripts to identify recurring themes and patterns in user feedback. We iteratively refined these codes through discussion until reaching consensus, then grouped them into higher-level themes. This analysis revealed primary themes, including task-dependent trust and adoption, balancing efficiency with authenticity, privacy and data ownership concerns, quality control and oversight, and implementation requirements.

\section{Results}

\begin{figure}[h!]
    \centering
    \includegraphics[width=\linewidth]{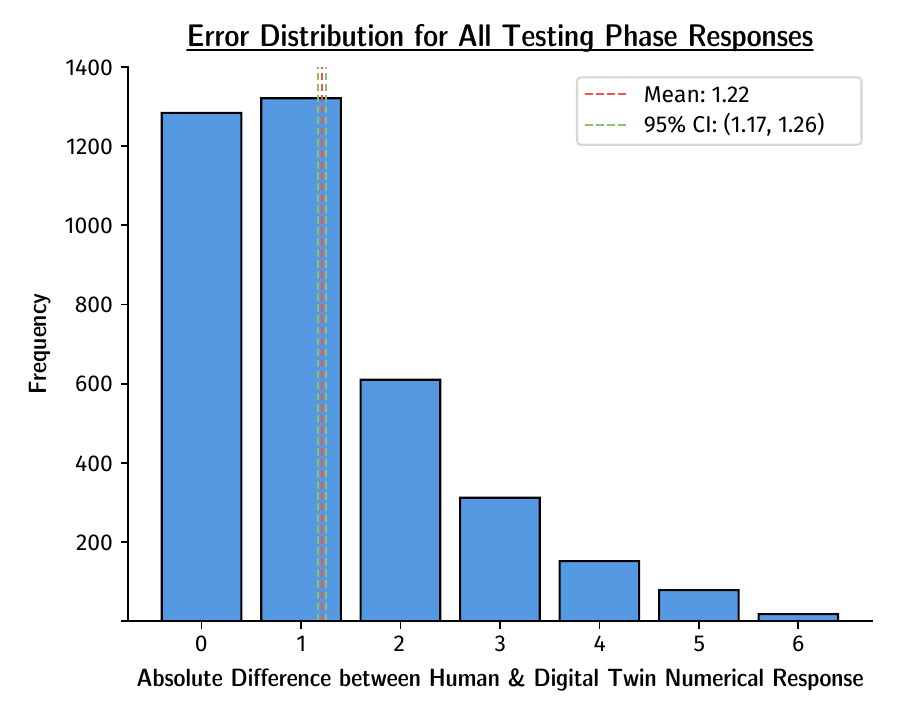}
    \Description[Absolute difference error for all responses]{Bar chart showing the distribution of the absolute difference error for all question responses from the testing phase. An absolute difference of 0 indicates identical responses between Human and Digital Twin on a Likert scale. Vertical lines represent the mean error and the 95\% confidence interval. Out of the 3,784 total questions, just under 1,300 were answered with 0 difference between the human and AI responses. Around 1300 saw an error of 1, a little over 600 saw an error of 2, and a small trail of questions were 3, 4, 5, or 6 points off. The mean was placed at 1.22, with the 95\% Confidence Interval spanning 1.17 and 1.26.}
    \caption{Distribution of absolute difference error for all test-phase responses (n = 3,784). An absolute difference of 0 means identical Human and Digital Twin Likert answers. Vertical lines show the mean error and 95\% confidence interval.}
    \label{fig:error1}
\end{figure}

\begin{figure}[h!]
    \centering
    \includegraphics[width=\linewidth]{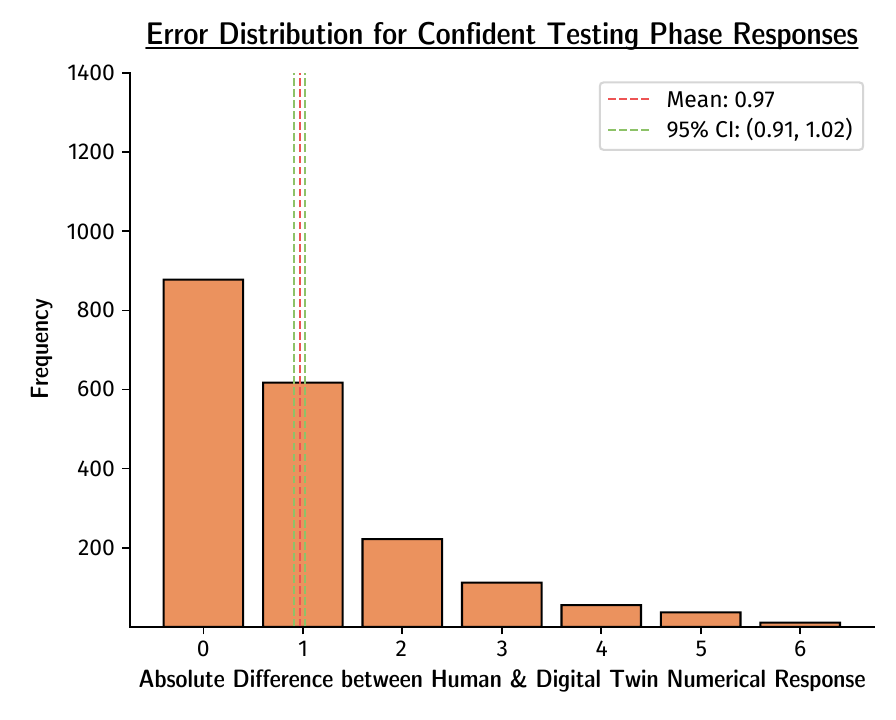}
    \Description[High-confidence absolute difference error]{Bar chart showing the distribution of the absolute difference error for responses where the Digital Twin displayed high confidence. High-confidence questions (>75\%) were answered automatically, with others deferred to human input. Vertical lines represent the mean error and the 95\% confidence interval. Out of the 1,933 questions, just under 900 were answered with 0 difference between the human and AI responses. Around 600 saw an error of 1, a little over 200 saw an error of 2, and a small trail of questions were 3, 4, 5, or 6 points off. The mean was placed at 0.97, with the 95\% Confidence Interval spanning 0.91 and 1.02.}
    \caption{Distribution of absolute difference error for test-phase responses where the digital twin had high confidence (>75\%) (n = 1,933). These questions were automatically answered, while others were deferred to humans. Vertical lines mark the mean error and 95\% confidence interval.}
    \label{fig:error2}
\end{figure}

\subsection{Performance Metrics}
        
\subsubsection{Accuracy of Digital Twin Predictions}
We measured the error of the digital twin's responses by computing the absolute difference between the digital twin's predictions and the participant's answers on a 7-point Likert scale. For questions the LLM answered independently, we compared its predictions with the participant's answers from the reflection survey. For questions the LLM deferred to the participant (it still made predictions in the background), we compared these predictions with the participant's actual answers. As shown in Figure~\ref{fig:error1}, the mean error across all testing phase responses was 1.22. As shown in Figure~\ref{fig:error2}, the mean error across all testing phase responses where the LLM answered in the user's place (i.e., it displayed high confidence) was 0.97. These results suggest that a digital twin is able to simulate an individual crowd worker's survey tasks with a high degree of accuracy.  

\subsubsection{Confidence of Digital Twin}
We measured the confidence of the digital twin for each response based on the token probability of the LLM's numerical Likert scale prediction. The mean confidence across all question responses was 70.36\%. Due to the 75\% confidence threshold in our system, the system accepted the digital twin answer for 51.1\% of questions on average across all testing phase responses and deferred the remaining 48.9\% to human participants. For additional quantitative results, see Appendix \ref{additional-results}.

\subsection{Crowd Worker Perspectives}

\subsubsection{Participants Recognize Accuracy but Emphasize the Need for Trust and Oversight}
Participants acknowledged the digital twin's accuracy, with 60.7\% of Prolific respondents reporting its recommendations often aligned with their choices and 31.5\% noting they sometimes aligned (see Table~\ref{tab:digital_twin_reflection_survey} for more details). Participants described the system as typically accurate, deviating by only one point (P2, P4). However, trust varied: only 5.7\% fully trusted it, while 25.0\% expressed no trust. P2 remarked, \textit{``If I worked with the same AI over and over, and it consistently chose things that were close to what I would choose, then I would probably trust it more.''} Human oversight was seen as essential, with 42.0\% of respondents preferring to monitor the system's decisions. P5 emphasized this, stating, \textit{``I'd like to see what AI has written... it can make mistakes.''}

\subsubsection{Participants Value Efficiency but Highlight the Need for Human Involvement in Meaningful Tasks}
The digital twin was praised for saving time, with 70.5\% of respondents citing task automation as a key benefit and 52.3\% valuing reduced decision fatigue. Participants appreciated AI for routine tasks and expressed frustration with repetitive, unstimulating work. P2 noted, \textit{``It's hard when I'm dealing with the same kind of basic economic questions over and over again... I've probably seen them a thousand times.''} Many favored a hybrid approach, allowing AI to handle routine tasks while humans focused on more engaging work. As P5 put it, \textit{``AI could help with the repetitive questions... but human input is still needed, especially when the questions require thinking or writing.''} 

\subsubsection{Participants Divided on Privacy, Prioritizing Authenticity Over Ethical Concerns}
Privacy opinions were mixed: 36.7\% of respondents were comfortable sharing data to improve accuracy, while 31.0\% were uncomfortable. Similarly, opinions on crowd workers owning and monetizing digital twins were split, with 37.5\% viewing it as harmful to privacy and 28.4\% seeing it as beneficial. Many accepted data sharing as part of ``life online,'' and prioritized authenticity over privacy (P2, P3, P4). As P3 explained, \textit{``My only concern is the capabilities of the LLM to truly represent me.''} Participants drew boundaries around sensitive data, such as personally identifiable information (PII), while appreciating AI's potential to enhance their work.

\subsubsection{Improvements in Accuracy, Transparency, and Hybrid Models Key to Adoption}
Key improvements included better prediction accuracy (67.0\%), greater transparency in data usage (65.9\%), and more control over privacy (56.8\%). The flexibility to toggle between AI and human input was also prioritized. P4 highlighted the hybrid model's appeal, noting, \textit{``If a study seems fun, let me figure this one out... workplace ones are dry, maybe a digital twin.''} P5 added, \textit{``If it's paying more, faster, and I can monitor it, I'd prefer the hybrid approach.''}

\subsubsection{Adoption and Recommendations Depend on Proven Benefits, Efficiency, and Human Oversight}
Interest in digital twins was mixed: 48.9\% expressed interest in future use, 28.4\% were unsure, and 22.8\% were unlikely to use the system. Participants emphasized the importance of proven benefits, such as improved accuracy (73.9\%), cost efficiency (54.5\%), and better privacy handling (50.0\%). P1 expressed reservations, saying, \textit{``I'm not comfortable with AI... but if it reduces scandalous activity, I'm fine with it.''} Others were optimistic about a hybrid approach, with P1 affirming, \textit{``hybrid is what's going to happen.''} Human oversight remained critical for trust and adaptability.

\subsection{Researcher Perspectives}

\subsubsection{Researchers Split on AI Detection and Trust in Diverse Crowd Work Tasks}
Researchers expressed mixed opinions on detecting AI in crowd work. P7 highlighted challenges in identifying AI use in emotional response surveys, while P6 expressed trust in participants, stating, \textit{``as a social scholar, I never thought about my respondents using AI.''} For heterogeneous, non-text-based tasks, P8 noted, \textit{``we cannot think of a way that you can easily adapt an AI to do the task.''}

\subsubsection{Researchers Prefer Human Responses but Acknowledge Cost Benefits of Human-AI Hybrid Approaches}Researchers strongly favored human responses for their authenticity and contextual richness. P9 stated, \textit{``ideally, I prefer entirely human responses despite the challenges, as they offer insights into contextual and temporal factors.''}, while P7 found hybrid systems \textit{``a preferable option.''} Others, like P8, viewed hybrid approaches as \textit{``the worst of both worlds,''} believing that most tasks that could reliably use digital twins could just use AI entirely. Despite this, AI's cost and scalability benefits were acknowledged, with some willing to adopt AI if it was \textit{``half the human cost''} (P6, P7). P9 observed, \textit{``A hybrid approach, like your digital twin, combines efficiency with personalization but requires ongoing human feedback to remain effective.''} 

\subsubsection{Transparency About AI Usage is Crucial for Research Integrity and Peer Review}Researchers stressed the importance of distinguishing AI from human responses for validity and peer review. P6 advocated for \textit{``clear information about whether a question is answered by [AI] or [human],''} allowing easier adjustments if AI usage is restricted. P9 emphasized transparency in the digital twin's training level for meaningful analysis, while P8 warned that excessive AI reliance could undermine research integrity. P6 also highlighted concerns about AI assumptions, noting \textit{``these AI tools may be based on assumptions that people from the same school will hold similar opinions.''} 

\subsubsection{Complex Tasks and Human Nuance Remain Beyond AI's Capabilities}
Researchers highlighted AI's limitations in handling nuanced tasks, such as complex social issues and morally ambiguous topics (P9). P8 noted, \textit{``It's difficult for AI to fully capture a person's complexity, as even a snapshot of someone is incomplete.''}  Variability in human opinions poses another challenge, with P8 emphasizing how responses shift based on phrasing and how digital twins may quickly become outdated as views evolve. P9 pointed out, \textit{``you want to sometimes know the things that are influencing people's responses at that moment.''} Additionally, AI often oversimplifies group representation, assuming, as P6 warned, \textit{``people from similar backgrounds hold similar opinions,''} ignoring individual variation. These challenges highlight AI's struggle to reflect human perspectives' diverse, dynamic nature.

\subsubsection{Clear Benchmarks and Standards Needed for Future AI Integration}
Researchers foresee greater AI use in crowd work but remain skeptical about full automation. P9 noted the digital twin system might work for binary tasks but warned, \textit{``for nuanced scales or exploratory research, small variations could matter,''} emphasizing the need to understand AI's limitations. P7 expressed concerns about unknowingly receiving predominantly AI-generated responses and stressed the importance of benchmarks, stating, \textit{``95\% accurate, for everyone, not 99\% for some and 89\% for others.''} P8 agreed, adding, \textit{``there must be a benchmark to which you compare the accuracy.''} While researchers remain cautious to protect research validity, they acknowledge AI's potential when supported by clear standards and benchmarks.

\section{Discussion}

Our study highlights key considerations for integrating AI in crowd work through digital twins. While our quantitative results show promising accuracy levels, researchers emphasized that numerical accuracy alone is insufficient, given the dynamic nature of human opinions and the influence of external events. This underscores the need for more sophisticated evaluation frameworks, such as routine sentiment checks or updates on personal opinions following significant social events.

We found a notable alignment between researcher and worker interests regarding data ownership and working conditions, with both groups emphasizing the importance of worker agency and control over digital representations. The effectiveness of digital twins appears highly task-dependent, with clear applications for repetitive elements like demographic questions while preserving human engagement for complex tasks requiring judgment. This natural division suggests opportunities for hybrid systems to improve working conditions and research integrity. Moving forward, aligning AI capabilities with worker needs and ethical considerations will be critical to ensuring that digital twins support sustainable and fair crowd work ecosystems.

Privacy and data usage are central to the ethical implementation of this study. To ensure transparency and participant awareness, all demographic information in this study was collected with informed consent and securely stored in compliance with human subjects research guidelines. While leveraging demographic data to tailor digital twins can enhance personalization, it also introduces risks related to data security and misuse; thus, strict safeguards were implemented to prevent unauthorized access and ensure data confidentiality. Inclusive demographic questions (see Appendix \ref{survey-questions}) were also used to promote fairness and recognition of individual diversity. Moving forward, research should prioritize methods that avoid reinforcing assumptions associated with demographic categories.

The ethical implications of integrating AI into crowd work also extend beyond data privacy to include concerns about the potential replacement of human judgment, along with its impact on worker livelihood and job security. While digital twins can alleviate repetitive tasks and reduce cognitive load, their widespread use may reduce demand for human input, affecting income stability for crowd workers. This emphasizes the need to deploy AI as a collaborative tool that enhances human capabilities rather than replacing them, preserving opportunities for meaningful and creative tasks that require human judgment.

\section{Limitations and Future Work}

From a technical perspective, our implementation using general-purpose LLMs (GPT-4o) presents both opportunities and limitations. The inherent biases in historical training data and the potential inability to capture responses to current events suggest the need for alternative approaches. However, resource constraints present practical challenges when training specialized models. Future work might explore creative solutions like enhanced prompting strategies or hybrid architectures that balance accuracy with resource efficiency.

Additionally, several limitations of our study warrant consideration, including sample size constraints and potential self-selection bias in our interview participants. Differences in research methodologies among participating researchers could also influence attitudes toward AI integration. For instance, one psychology researcher expressed little concern about AI use (P8), given that their research relied on non-text-based survey designs. Future research should investigate digital twins' applicability to tasks involving free-text responses, subjective assessments, and contextual interpretation, exploring how AI can effectively support diverse research needs.

Finally, the study's focus on Likert-scale tasks limits the generalizability of our findings to more complex or nuanced annotation work. Future research should focus on expanding digital twin capabilities beyond Likert scales, developing flexible systems for AI control, implementing adaptive oversight mechanisms, and making crowd work more accessible to diverse populations. While digital twins demonstrate promise to enhance the efficiency and quality of crowd work, their successful implementation depends on balancing technical capabilities with the evolving needs of researchers and preserving the integrity of human-centered research.

\section{Conclusion}

Our study highlights the potential of hybrid AI-human systems, particularly digital twins, to revolutionize crowd work by balancing efficiency and authenticity. While the proposed framework automates repetitive tasks and maintains research integrity, it unders cores the importance of human agency and task-specific AI deployment. Both crowd workers and researchers acknowledge the benefits of such systems but emphasize the need for transparency, trust, and oversight. Our findings demonstrate that aligning worker and researcher interests through clear benchmarks, flexible AI usage, and robust privacy safeguards can lead to ethical and inclusive AI integration. Future work should expand on these foundations, addressing limitations in model adaptability, task diversity, and system scalability to ensure broader applicability and fairness in AI-augmented research.


\bibliographystyle{ACM-Reference-Format}
\bibliography{refs}


\begin{thebibliography}{33}


\ifx \showCODEN    \undefined \def \showCODEN     #1{\unskip}     \fi
\ifx \showDOI      \undefined \def \showDOI       #1{#1}\fi
\ifx \showISBNx    \undefined \def \showISBNx     #1{\unskip}     \fi
\ifx \showISBNxiii \undefined \def \showISBNxiii  #1{\unskip}     \fi
\ifx \showISSN     \undefined \def \showISSN      #1{\unskip}     \fi
\ifx \showLCCN     \undefined \def \showLCCN      #1{\unskip}     \fi
\ifx \shownote     \undefined \def \shownote      #1{#1}          \fi
\ifx \showarticletitle \undefined \def \showarticletitle #1{#1}   \fi
\ifx \showURL      \undefined \def \showURL       {\relax}        \fi
\providecommand\bibfield[2]{#2}
\providecommand\bibinfo[2]{#2}
\providecommand\natexlab[1]{#1}
\providecommand\showeprint[2][]{arXiv:#2}

\bibitem[{AIAA}(2020)]%
        {aiaa_digital_2020}
\bibfield{author}{\bibinfo{person}{{AIAA}}.} \bibinfo{year}{2020}\natexlab{}.
\newblock \bibinfo{title}{Digital {Twin}: {Definition} \& {Value} – {An} {AIAA} and {AIA} {Position} {Paper}}.
\newblock
\newblock
\urldef\tempurl%
\url{https://www.aia-aerospace.org/publications/digital-twin-definition-value-an-aiaa-and-aia-position-paper/}
\showURL{%
\tempurl}


\bibitem[Argyle et~al\mbox{.}(2022)]%
        {argyle_out_2022}
\bibfield{author}{\bibinfo{person}{Lisa~P. Argyle}, \bibinfo{person}{Ethan~C. Busby}, \bibinfo{person}{Nancy Fulda}, \bibinfo{person}{Joshua Gubler}, \bibinfo{person}{Christopher Rytting}, {and} \bibinfo{person}{David Wingate}.} \bibinfo{year}{2022}\natexlab{}.
\newblock \bibinfo{title}{Out of {One}, {Many}: {Using} {Language} {Models} to {Simulate} {Human} {Samples}}.
\newblock
\newblock
\urldef\tempurl%
\url{https://doi.org/10.1017/pan.2023.2}
\showDOI{\tempurl}


\bibitem[Babanagar et~al\mbox{.}(2025)]%
        {babanagar_digital_2025}
\bibfield{author}{\bibinfo{person}{Nandeesh Babanagar}, \bibinfo{person}{Brian Sheil}, \bibinfo{person}{Jelena Ninić}, \bibinfo{person}{Qianbing Zhang}, {and} \bibinfo{person}{Stuart Hardy}.} \bibinfo{year}{2025}\natexlab{}.
\newblock \showarticletitle{Digital twins for urban underground space}.
\newblock \bibinfo{journal}{\emph{Tunnelling and Underground Space Technology}}  \bibinfo{volume}{155} (\bibinfo{date}{Jan.} \bibinfo{year}{2025}), \bibinfo{pages}{106140}.
\newblock
\showISSN{0886-7798}
\urldef\tempurl%
\url{https://doi.org/10.1016/j.tust.2024.106140}
\showDOI{\tempurl}


\bibitem[{Daniel Silverman} et~al\mbox{.}(2020)]%
        {daniel_silverman_can_2020}
\bibfield{author}{\bibinfo{person}{{Daniel Silverman}}, \bibinfo{person}{{Daniel Kent}}, {and} \bibinfo{person}{{Christopher Gelpi}}.} \bibinfo{year}{2020}\natexlab{}.
\newblock \showarticletitle{Can {Factual} {Misperceptions} be {Corrected}? {An} {Experiment} on {American} {Public} {Fears} of {Terrorism}}.
\newblock  (\bibinfo{date}{June} \bibinfo{year}{2020}).
\newblock
\urldef\tempurl%
\url{https://osf.io/a7uk3/}
\showURL{%
\tempurl}
\newblock
\shownote{Publisher: OSF}.


\bibitem[Davern et~al\mbox{.}(2024)]%
        {davern_general_2024}
\bibfield{author}{\bibinfo{person}{Michael Davern}, \bibinfo{person}{Rene Bautista}, \bibinfo{person}{Jeremy Freese}, \bibinfo{person}{Pamela Herd}, {and} \bibinfo{person}{Stephen~L. Morgan}.} \bibinfo{year}{2024}\natexlab{}.
\newblock \bibinfo{title}{General {Social} {Survey} 1972-2024}.
\newblock
\newblock
\urldef\tempurl%
\url{gssdataexplorer.norc.org}
\showURL{%
\tempurl}


\bibitem[Drăghici et~al\mbox{.}(2023)]%
        {draghici_revolutionizing_2023}
\bibfield{author}{\bibinfo{person}{Diana-Elena Drăghici}, \bibinfo{person}{Andreea Orîndaru}, \bibinfo{person}{Mihaela Constantinescu}, {and} \bibinfo{person}{Alina Zelezneac}.} \bibinfo{year}{2023}\natexlab{}.
\newblock \showarticletitle{Revolutionizing {Marketing} {Research} {Through} {AI}: comprehensive review of the past, present, and future}.
\newblock \bibinfo{journal}{\emph{Journal of Emerging Trends in Marketing and Management}} \bibinfo{volume}{I}, \bibinfo{number}{1} (\bibinfo{date}{May} \bibinfo{year}{2023}), \bibinfo{pages}{39--45}.
\newblock
\showISSN{2537-5865}
\urldef\tempurl%
\url{http://www.etimm.ase.ro/RePEc/aes/jetimm/2023/ETIMM_V01_2023_73.pdf}
\showURL{%
\tempurl}
\newblock
\shownote{Publisher: The Bucharest University of Economic Studies Publishing House}.


\bibitem[Gilardi et~al\mbox{.}(2023)]%
        {gilardi2023chatgpt}
\bibfield{author}{\bibinfo{person}{Fabrizio Gilardi}, \bibinfo{person}{Meysam Alizadeh}, {and} \bibinfo{person}{Ma{\"e}l Kubli}.} \bibinfo{year}{2023}\natexlab{}.
\newblock \showarticletitle{ChatGPT outperforms crowd workers for text-annotation tasks}.
\newblock \bibinfo{journal}{\emph{Proceedings of the National Academy of Sciences}} \bibinfo{volume}{120}, \bibinfo{number}{30} (\bibinfo{year}{2023}), \bibinfo{pages}{e2305016120}.
\newblock


\bibitem[Gray and Suri(2019)]%
        {gray2019ghost}
\bibfield{author}{\bibinfo{person}{Mary~L Gray} {and} \bibinfo{person}{Siddharth Suri}.} \bibinfo{year}{2019}\natexlab{}.
\newblock \bibinfo{booktitle}{\emph{Ghost work: How to stop Silicon Valley from building a new global underclass}}.
\newblock \bibinfo{publisher}{Eamon Dolan Books}.
\newblock


\bibitem[Hewitt et~al\mbox{.}(2024)]%
        {hewitt_predicting_2024}
\bibfield{author}{\bibinfo{person}{Luke Hewitt}, \bibinfo{person}{Ashwini Ashokkumar}, \bibinfo{person}{Isaias Ghezae}, {and} \bibinfo{person}{Robb Willer}.} \bibinfo{year}{2024}\natexlab{}.
\newblock \showarticletitle{Predicting {Results} of {Social} {Science} {Experiments} {Using} {Large} {Language} {Models}}.
\newblock  (\bibinfo{date}{Aug.} \bibinfo{year}{2024}).
\newblock


\bibitem[Huang et~al\mbox{.}(2024)]%
        {huang_simulation_2024}
\bibfield{author}{\bibinfo{person}{Yijun Huang}, \bibinfo{person}{Jihan Zhang}, \bibinfo{person}{Xi Chen}, \bibinfo{person}{Alan H.~F. Lam}, {and} \bibinfo{person}{Ben~M. Chen}.} \bibinfo{year}{2024}\natexlab{}.
\newblock \showarticletitle{From {Simulation} to {Prediction}: {Enhancing} {Digital} {Twins} with {Advanced} {Generative} {AI} {Technologies}}. In \bibinfo{booktitle}{\emph{2024 {IEEE} 18th {International} {Conference} on {Control} \& {Automation} ({ICCA})}}. \bibinfo{pages}{490--495}.
\newblock
\urldef\tempurl%
\url{https://doi.org/10.1109/ICCA62789.2024.10591881}
\showDOI{\tempurl}
\newblock
\shownote{ISSN: 1948-3457}.


\bibitem[Hughes et~al\mbox{.}(2022)]%
        {hughes_guidance_2022}
\bibfield{author}{\bibinfo{person}{Jennifer~L. Hughes}, \bibinfo{person}{Abigail~A. Camden}, \bibinfo{person}{Tenzin Yangchen}, \bibinfo{person}{Gabrielle P.~A. Smith}, \bibinfo{person}{Melanie~M. Domenech~Rodríguez}, \bibinfo{person}{Steven~V. Rouse}, \bibinfo{person}{C.~Peeper McDonald}, {and} \bibinfo{person}{Stella Lopez}.} \bibinfo{year}{2022}\natexlab{}.
\newblock \showarticletitle{Guidance for {Researchers} {When} {Using} {Inclusive} {Demographic} {Questions} for {Surveys}: {Improved} and {Updated} {Questions}}.
\newblock \bibinfo{journal}{\emph{Psi Chi Journal of Psychological Research}} \bibinfo{volume}{27}, \bibinfo{number}{4} (\bibinfo{year}{2022}), \bibinfo{pages}{232--255}.
\newblock
\showISSN{23257342}
\urldef\tempurl%
\url{https://doi.org/10.24839/2325-7342.JN27.4.232}
\showDOI{\tempurl}


\bibitem[IBM(2021)]%
        {ibm_dt}
\bibfield{author}{\bibinfo{person}{IBM}.} \bibinfo{year}{2021}\natexlab{}.
\newblock \bibinfo{title}{What {Is} a {Digital} {Twin}?}
\newblock
\newblock
\urldef\tempurl%
\url{https://www.ibm.com/topics/what-is-a-digital-twin}
\showURL{%
\tempurl}


\bibitem[Insider(2024)]%
        {businessinsider2024synthetic}
\bibfield{author}{\bibinfo{person}{Business Insider}.} \bibinfo{year}{2024}\natexlab{}.
\newblock \showarticletitle{The AI world's most valuable resource is running out, and it's scrambling to find an alternative: 'fake' data}.
\newblock  (\bibinfo{year}{2024}).
\newblock
\urldef\tempurl%
\url{https://www.businessinsider.com/ai-synthetic-data-industry-debate-over-fake-2024-8}
\showURL{%
\tempurl}
\newblock
\shownote{Accessed: 2025-01-19}.


\bibitem[Kanda et~al\mbox{.}(2022)]%
        {kanda_efficient_2022}
\bibfield{author}{\bibinfo{person}{Tomoya Kanda}, \bibinfo{person}{Hiroyoshi Ito}, {and} \bibinfo{person}{Atsuyuki Morishima}.} \bibinfo{year}{2022}\natexlab{}.
\newblock \showarticletitle{Efficient {Evaluation} of {AI} {Workers} for the {Human}+{AI} {Crowd} {Task} {Assignment}}. In \bibinfo{booktitle}{\emph{2022 {IEEE} {International} {Conference} on {Big} {Data} ({Big} {Data})}}. \bibinfo{pages}{3995--4001}.
\newblock
\urldef\tempurl%
\url{https://doi.org/10.1109/BigData55660.2022.10020844}
\showDOI{\tempurl}


\bibitem[Karya({[n.\,d.]})]%
        {karya}
\bibfield{author}{\bibinfo{person}{Karya}.} \bibinfo{year}{[n.\,d.]}\natexlab{}.
\newblock \bibinfo{title}{Karya Institute}.
\newblock
\newblock
\urldef\tempurl%
\url{https://institute.karya.in/}
\showURL{%
\tempurl}


\bibitem[{Katherine McCabe}(2019)]%
        {katherine_mccabe_public_2019}
\bibfield{author}{\bibinfo{person}{{Katherine McCabe}}.} \bibinfo{year}{2019}\natexlab{}.
\newblock \showarticletitle{Public {Opinion} and {Attributions} for {Health} {Care} {Costs}}.
\newblock  (\bibinfo{date}{May} \bibinfo{year}{2019}).
\newblock
\urldef\tempurl%
\url{https://osf.io/3pcdm/}
\showURL{%
\tempurl}
\newblock
\shownote{Publisher: OSF}.


\bibitem[Kobayashi et~al\mbox{.}(2021)]%
        {kobayashi_humanai_2021}
\bibfield{author}{\bibinfo{person}{Masaki Kobayashi}, \bibinfo{person}{Kei Wakabayashi}, {and} \bibinfo{person}{Atsuyuki Morishima}.} \bibinfo{year}{2021}\natexlab{}.
\newblock \showarticletitle{Human+{AI} {Crowd} {Task} {Assignment} {Considering} {Result} {Quality} {Requirements}}.
\newblock \bibinfo{journal}{\emph{Proceedings of the AAAI Conference on Human Computation and Crowdsourcing}}  \bibinfo{volume}{9} (\bibinfo{date}{Oct.} \bibinfo{year}{2021}), \bibinfo{pages}{97--107}.
\newblock
\showISSN{2769-1349}
\urldef\tempurl%
\url{https://doi.org/10.1609/hcomp.v9i1.18943}
\showDOI{\tempurl}


\bibitem[{Maureen Craig}(2017)]%
        {maureen_craig_racial_2017}
\bibfield{author}{\bibinfo{person}{{Maureen Craig}}.} \bibinfo{year}{2017}\natexlab{}.
\newblock \showarticletitle{Racial {Majority} \& {Minority} {Group} {Members}' {Psychological} and {Political} {Reactions} to {Minority} {Population} {Growth}}.
\newblock  (\bibinfo{date}{Nov.} \bibinfo{year}{2017}).
\newblock
\urldef\tempurl%
\url{https://osf.io/sazxn/}
\showURL{%
\tempurl}
\newblock
\shownote{Publisher: OSF}.


\bibitem[Nagaraj et~al\mbox{.}(2023)]%
        {nagaraj_augmenting_2023}
\bibfield{author}{\bibinfo{person}{Divya Nagaraj}, \bibinfo{person}{Priya Khandelwal}, \bibinfo{person}{Sandra Steyaert}, {and} \bibinfo{person}{Olivier Gevaert}.} \bibinfo{year}{2023}\natexlab{}.
\newblock \showarticletitle{Augmenting digital twins with federated learning in medicine}.
\newblock \bibinfo{journal}{\emph{The Lancet. Digital Health}} \bibinfo{volume}{5}, \bibinfo{number}{5} (\bibinfo{date}{May} \bibinfo{year}{2023}), \bibinfo{pages}{e251--e253}.
\newblock
\showISSN{2589-7500}
\urldef\tempurl%
\url{https://doi.org/10.1016/S2589-7500(23)00044-4}
\showDOI{\tempurl}


\bibitem[Padmakumar and He(2024)]%
        {padmakumar2024doeswritinglanguagemodels}
\bibfield{author}{\bibinfo{person}{Vishakh Padmakumar} {and} \bibinfo{person}{He He}.} \bibinfo{year}{2024}\natexlab{}.
\newblock \bibinfo{title}{Does Writing with Language Models Reduce Content Diversity?}
\newblock
\newblock
\showeprint[arxiv]{2309.05196}~[cs.CL]
\urldef\tempurl%
\url{https://arxiv.org/abs/2309.05196}
\showURL{%
\tempurl}


\bibitem[Park et~al\mbox{.}(2023)]%
        {park_generative_2023}
\bibfield{author}{\bibinfo{person}{Joon~Sung Park}, \bibinfo{person}{Joseph O'Brien}, \bibinfo{person}{Carrie~Jun Cai}, \bibinfo{person}{Meredith~Ringel Morris}, \bibinfo{person}{Percy Liang}, {and} \bibinfo{person}{Michael~S. Bernstein}.} \bibinfo{year}{2023}\natexlab{}.
\newblock \showarticletitle{Generative {Agents}: {Interactive} {Simulacra} of {Human} {Behavior}}. In \bibinfo{booktitle}{\emph{Proceedings of the 36th {Annual} {ACM} {Symposium} on {User} {Interface} {Software} and {Technology}}} \emph{(\bibinfo{series}{{UIST} '23})}. \bibinfo{publisher}{Association for Computing Machinery}, \bibinfo{address}{New York, NY, USA}, \bibinfo{pages}{1--22}.
\newblock
\showISBNx{9798400701320}
\urldef\tempurl%
\url{https://doi.org/10.1145/3586183.3606763}
\showDOI{\tempurl}


\bibitem[Park et~al\mbox{.}(2024)]%
        {park_generative_2024}
\bibfield{author}{\bibinfo{person}{Joon~Sung Park}, \bibinfo{person}{Carolyn~Q. Zou}, \bibinfo{person}{Aaron Shaw}, \bibinfo{person}{Benjamin~Mako Hill}, \bibinfo{person}{Carrie Cai}, \bibinfo{person}{Meredith~Ringel Morris}, \bibinfo{person}{Robb Willer}, \bibinfo{person}{Percy Liang}, {and} \bibinfo{person}{Michael~S. Bernstein}.} \bibinfo{year}{2024}\natexlab{}.
\newblock \bibinfo{title}{Generative {Agent} {Simulations} of 1,000 {People}}.
\newblock
\newblock
\urldef\tempurl%
\url{https://doi.org/10.48550/arXiv.2411.10109}
\showDOI{\tempurl}
\newblock
\shownote{arXiv:2411.10109 [cs]}.


\bibitem[Peffley and Hurwitz(2007)]%
        {peffley_persuasion_2007}
\bibfield{author}{\bibinfo{person}{Mark Peffley} {and} \bibinfo{person}{Jon Hurwitz}.} \bibinfo{year}{2007}\natexlab{}.
\newblock \showarticletitle{Persuasion and {Resistance}: {Race} and the {Death} {Penalty} in {America}}.
\newblock \bibinfo{journal}{\emph{American Journal of Political Science}} \bibinfo{volume}{51}, \bibinfo{number}{4} (\bibinfo{year}{2007}), \bibinfo{pages}{996--1012}.
\newblock
\showISSN{1540-5907}
\urldef\tempurl%
\url{https://doi.org/10.1111/j.1540-5907.2007.00293.x}
\showDOI{\tempurl}
\newblock
\shownote{\_eprint: https://onlinelibrary.wiley.com/doi/pdf/10.1111/j.1540-5907.2007.00293.x}.


\bibitem[Qloo({[n.\,d.]})]%
        {qloo}
\bibfield{author}{\bibinfo{person}{Qloo}.} \bibinfo{year}{[n.\,d.]}\natexlab{}.
\newblock \bibinfo{title}{Qloo {\textbar} Develop Personalized Experiences With Taste {AI}}.
\newblock
\newblock
\urldef\tempurl%
\url{https://www.qloo.com}
\showURL{%
\tempurl}


\bibitem[{Rebecca Bucci}(2023)]%
        {rebecca_bucci_accounting_2023}
\bibfield{author}{\bibinfo{person}{{Rebecca Bucci}}.} \bibinfo{year}{2023}\natexlab{}.
\newblock \bibinfo{title}{Accounting for the {Correlation} between {Perceived} {Risks} and {Rewards} to {Crime}}.
\newblock
\newblock
\urldef\tempurl%
\url{https://osf.io/4fsg5/}
\showURL{%
\tempurl}


\bibitem[{Rochelle Terman}(2020)]%
        {rochelle_terman_human_2020}
\bibfield{author}{\bibinfo{person}{{Rochelle Terman}}.} \bibinfo{year}{2020}\natexlab{}.
\newblock \showarticletitle{Human {Rights} {Shaming}, {Compliance}, and {Nationalist} {Backlash}}.
\newblock  (\bibinfo{date}{Jan.} \bibinfo{year}{2020}).
\newblock
\urldef\tempurl%
\url{https://osf.io/q8ra3/}
\showURL{%
\tempurl}
\newblock
\shownote{Publisher: OSF}.


\bibitem[Salganik(2019)]%
        {salganik2019bit}
\bibfield{author}{\bibinfo{person}{Matthew~J Salganik}.} \bibinfo{year}{2019}\natexlab{}.
\newblock \bibinfo{booktitle}{\emph{Bit by bit: Social research in the digital age}}.
\newblock \bibinfo{publisher}{Princeton University Press}.
\newblock


\bibitem[Veselovsky et~al\mbox{.}(2023b)]%
        {veselovsky_prevalence_2023}
\bibfield{author}{\bibinfo{person}{Veniamin Veselovsky}, \bibinfo{person}{Manoel~Horta Ribeiro}, \bibinfo{person}{Philip Cozzolino}, \bibinfo{person}{Andrew Gordon}, \bibinfo{person}{David Rothschild}, {and} \bibinfo{person}{Robert West}.} \bibinfo{year}{2023}\natexlab{b}.
\newblock \bibinfo{title}{Prevalence and prevention of large language model use in crowd work}.
\newblock
\newblock
\urldef\tempurl%
\url{https://doi.org/10.48550/arXiv.2310.15683}
\showDOI{\tempurl}
\newblock
\shownote{arXiv:2310.15683 [cs]}.


\bibitem[Veselovsky et~al\mbox{.}(2023a)]%
        {veselovsky_artificial_2023}
\bibfield{author}{\bibinfo{person}{Veniamin Veselovsky}, \bibinfo{person}{Manoel~Horta Ribeiro}, {and} \bibinfo{person}{Robert West}.} \bibinfo{year}{2023}\natexlab{a}.
\newblock \bibinfo{title}{Artificial {Artificial} {Artificial} {Intelligence}: {Crowd} {Workers} {Widely} {Use} {Large} {Language} {Models} for {Text} {Production} {Tasks}}.
\newblock
\newblock
\urldef\tempurl%
\url{https://arxiv.org/abs/2306.07899v1}
\showURL{%
\tempurl}


\bibitem[Wang et~al\mbox{.}(2024b)]%
        {wang2024large}
\bibfield{author}{\bibinfo{person}{Angelina Wang}, \bibinfo{person}{Jamie Morgenstern}, {and} \bibinfo{person}{John~P Dickerson}.} \bibinfo{year}{2024}\natexlab{b}.
\newblock \showarticletitle{Large language models cannot replace human participants because they cannot portray identity groups}.
\newblock \bibinfo{journal}{\emph{arXiv preprint arXiv:2402.01908}} (\bibinfo{year}{2024}).
\newblock


\bibitem[Wang et~al\mbox{.}(2024a)]%
        {wang_human-llm_2024}
\bibfield{author}{\bibinfo{person}{Xinru Wang}, \bibinfo{person}{Hannah Kim}, \bibinfo{person}{Sajjadur Rahman}, \bibinfo{person}{Kushan Mitra}, {and} \bibinfo{person}{Zhengjie Miao}.} \bibinfo{year}{2024}\natexlab{a}.
\newblock \showarticletitle{Human-{LLM} Collaborative Annotation Through Effective Verification of {LLM} Labels}. In \bibinfo{booktitle}{\emph{Proceedings of the {CHI} Conference on Human Factors in Computing Systems}} (Honolulu {HI} {USA}, 2024-05-11). \bibinfo{publisher}{{ACM}}, \bibinfo{pages}{1--21}.
\newblock
\showISBNx{9798400703300}
\urldef\tempurl%
\url{https://doi.org/10.1145/3613904.3641960}
\showDOI{\tempurl}


\bibitem[Wang et~al\mbox{.}(2022)]%
        {wang_interpretable_2022}
\bibfield{author}{\bibinfo{person}{Yunlong Wang}, \bibinfo{person}{Priyadarshini Venkatesh}, {and} \bibinfo{person}{Brian~Y Lim}.} \bibinfo{year}{2022}\natexlab{}.
\newblock \showarticletitle{Interpretable {Directed} {Diversity}: {Leveraging} {Model} {Explanations} for {Iterative} {Crowd} {Ideation}}. In \bibinfo{booktitle}{\emph{Proceedings of the 2022 {CHI} {Conference} on {Human} {Factors} in {Computing} {Systems}}} \emph{(\bibinfo{series}{{CHI} '22})}. \bibinfo{publisher}{Association for Computing Machinery}, \bibinfo{address}{New York, NY, USA}, \bibinfo{pages}{1--28}.
\newblock
\showISBNx{978-1-4503-9157-3}
\urldef\tempurl%
\url{https://doi.org/10.1145/3491102.3517551}
\showDOI{\tempurl}


\bibitem[Wu et~al\mbox{.}(2023)]%
        {wu2023llms}
\bibfield{author}{\bibinfo{person}{Tongshuang Wu}, \bibinfo{person}{Haiyi Zhu}, \bibinfo{person}{Maya Albayrak}, \bibinfo{person}{Alexis Axon}, \bibinfo{person}{Amanda Bertsch}, \bibinfo{person}{Wenxing Deng}, \bibinfo{person}{Ziqi Ding}, \bibinfo{person}{Bill Guo}, \bibinfo{person}{Sireesh Gururaja}, \bibinfo{person}{Tzu-Sheng Kuo}, {et~al\mbox{.}}} \bibinfo{year}{2023}\natexlab{}.
\newblock \showarticletitle{Llms as workers in human-computational algorithms? replicating crowdsourcing pipelines with llms}.
\newblock \bibinfo{journal}{\emph{arXiv preprint arXiv:2307.10168}} (\bibinfo{year}{2023}).
\newblock


\end{thebibliography}
\pagebreak

\appendix

\section{Additional Results}
\label{additional-results}

\begin{table*}[htb]
\resizebox{\textwidth}{!}{%
\begin{tabular}{@{}l|l|l|l|l@{}}
\toprule
\textbf{Testing Results} &
  \multicolumn{1}{c}{\textbf{Mean Absolute Difference (95\% CI)}} &
  \multicolumn{1}{c}{\textbf{Mean Confidence (\%) (95\% CI)}} &
  \multicolumn{1}{c}{\textbf{Total Number of Responses}} \\ \midrule
\textbf{All Testing Responses} &
  1.22 (1.17–1.26) &
  70.36 (69.62–71.09) &
  3,776 \\ \midrule
\textbf{\begin{tabular}[c]{@{}l@{}}All Testing Responses \\ Independently Answered by Digital Twin\end{tabular}} &
  0.97 (0.91–1.02) &
  88.82 (88.47–89.16) &
  \begin{tabular}[c]{@{}l@{}}1,931 \\ (51.1\%)\end{tabular} \\ \midrule
\textbf{\begin{tabular}[c]{@{}l@{}}All Testing Responses \\ Deferred and Answered by Human\end{tabular}} &
  1.47 (1.42–1.53) &
  51.03 (50.24–51.82) &
  \begin{tabular}[c]{@{}l@{}}1,845 \\ (48.9\%)\end{tabular} \\ \bottomrule
\end{tabular}%
}
\caption{Testing results with mean absolute difference, mean confidence, and total number of responses.}
\label{tab:testing-results}
\end{table*}

\begin{table*}[htb]
\centering
\renewcommand{\arraystretch}{1.2} 
\resizebox{\textwidth}{!}{%
\begin{tabular}{@{}p{5cm}|p{2.5cm}|p{2.5cm}|p{2.5cm}|p{2.5cm}|p{2.5cm}@{}}
{\textbf{Question}} & 
{\textbf{Option 1}} & 
{\textbf{Option 2}} & 
{\textbf{Option 3}} & 
{\textbf{Option 4}} & 
{\textbf{Option 5}} \\ 
\hline
\textbf{How well do the recommendations from the digital twin align with your choices?} &
1.1\% Always align &
60.7\% Often align &
31.5\% Sometimes align &
6.7\% Rarely align &
0.0\% Never align \\ \midrule
\textbf{Do you think systems like this could improve or harm the conditions for crowd workers?} &
9.1\% Strongly improve &
34.1\% Somewhat improve &
19.3\% No significant impact &
27.3\% Somewhat harm &
10.2\% Strongly harm \\ \midrule
\textbf{How do you feel about the concept of using digital twins for survey-answering tasks?} &
23.0\% Very excited &
28.7\% Somewhat excited &
18.4\% Neutral &
16.1\% Somewhat skeptical &
13.8\% Very skeptical \\ \midrule
\textbf{How comfortable are you with sharing personal data to improve the accuracy of your digital twin?} &
19.5\% Very comfortable &
17.2\% Somewhat comfortable &
19.5\% Neutral &
17.2\% Somewhat uncomfortable &
13.8\% Very uncomfortable \\ \midrule
\textbf{How would allowing crowd workers to own and monetize their digital twins affect your data privacy concerns?} &
9.1\% Strongly improve &
19.3\% Somewhat improve &
34.1\% No significant impact &
29.5\% Somewhat harm &
8.0\% Strongly harm \\ \midrule
\textbf{Would you accept slightly lower pay if the platform reduced your workload by using AI to handle repetitive tasks?} &
10.2\% Yes, definitely &
21.6\% Yes, to some extent &
19.3\% Not sure &
20.5\% No, unlikely &
20.5\% No, definitely not \\ \midrule
\textbf{To what extent would you trust a digital twin to make decisions on your behalf?} &
5.7\% Fully trust it to make decisions for me &
42.0\% Trust it with some decisions, but prefer oversight &
27.3\% Trust it only with minor decisions &
25.0\% Don't trust it to make any decisions &
N/A \\ \midrule
\textbf{Out of everything in this system, what stood out to you the most?} &
33.0\% Ease of use &
31.8\% Alignment with my choices &
11.4\% Ethical and privacy concerns &
15.9\% Accuracy of predictions &
5.7\% Potential benefits for crowd workers \\ \midrule
\textbf{In what areas has (or do you expect) the digital twin to help you the most? (select all that apply)} &
70.5\% Saving time by automating tasks &
52.3\% Reducing decision fatigue by providing recommendations &
52.3\% Gaining insights about personal habits or behavior &
23.9\% Improving my physical or mental well-being &
N/A \\ \midrule
\textbf{What improvements would make you more likely to use a digital twin regularly? (select all that apply)} &
67.0\% Better accuracy of predictions &
65.9\% More transparency about how data is used &
56.8\% Greater control over data collection and privacy &
37.5\% Improved user interface and ease of use &
42.0\% Stronger alignment with my personal values \\ \midrule
\textbf{Would you use a digital twin like this in the future?} &
14.8\% Yes, definitely &
34.1\% Yes, likely &
28.4\% Maybe, unsure &
14.8\% No, unlikely &
8.0\% No, definitely not \\ \midrule
\textbf{Would you recommend this system to others in your field?} &
18.4\% Yes, highly recommend &
20.7\% Yes, recommend with some reservations &
32.2\% Neutral &
19.5\% No, would not recommend &
9.2\% No, strongly against recommending \\ \midrule
\textbf{What would increase your confidence in adopting or recommending this platform? (select all that apply)} &
42.0\% Clearer explanation of how it works &
54.5\% Demonstrated cost and time efficiency &
73.8\% Evidence of improved accuracy over alternatives &
50.0\% Better handling of ethical or privacy concerns &
N/A \\ \bottomrule
\end{tabular}%
}
\caption{Summary of Reflection Survey Responses}
\label{tab:digital_twin_reflection_survey}
\end{table*}

\begin{figure}[htbp]
    \centering
    \includegraphics[width=\linewidth]{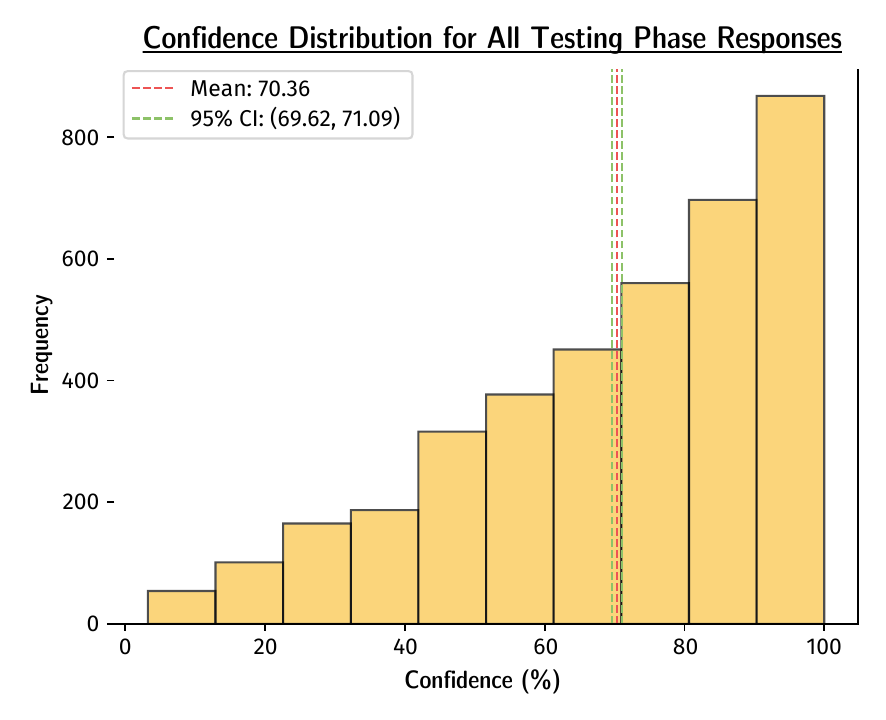}
    \Description[Confidence distribution for testing phase responses.]{Histogram showing the distribution of the Digital Twin's confidence levels for all question responses from the testing phase. The x-axis represents confidence percentages (0\% to 100\%), and the y-axis represents frequency. The mean confidence is marked at 70.36\%, with a 95\% confidence interval ranging from 69.62\% to 71.09\%. Confidence levels increase toward 100\%, with higher frequencies observed at higher confidence percentages.}
    \caption{The distribution of digital twin confidence for all question responses from the testing phase.}
    \label{fig:confidence}
\end{figure}

\section{Interview Protocol}
\begin{enumerate}
    \item \textbf{Introduction and Consent (5 minutes):}
    \begin{itemize}
        \item Researchers introduce themselves and explain the purpose of the study.
        \item Participants confirm they have read and signed the consent form.
    \end{itemize}
    
    \item \textbf{Prototype Use (20 minutes):}
        \begin{itemize}
            \item Participants use our prototype system by answering a series of surveys and interacting with a ``digital twin'' that assists them in answering questions.
            \item Participants evaluate the digital twin based on perceived accuracy and alignment with their real preferences. They also answer feedback questions about the system.
        \end{itemize}
    \item \textbf{Discussion (35 minutes):} 
    \begin{itemize}
        \item Questions focus on the participants' thoughts on integrating digital twins into their workflows, their perceptions of the hybrid approach's validity, and potential barriers to adoption.
    \end{itemize}
    \item \textbf{Conclusion (5 minutes):}
    \begin{itemize}
        \item Summarize key discussion points and ask if participants have additional thoughts.
        \item Thank participants for their time.
    \end{itemize}
\end{enumerate}

\section {Survey Questions}
\label{survey-questions}

The following list provides a breakdown of the sources referenced for each survey question.  The ``Supplementary Materials'' document includes the complete list of survey questions included in the study. Many of these questions are either directly taken from or inspired by established research studies, as detailed below. The sources were identified through the work of Hewitt and colleagues, who compiled 70 nationally representative U.S. studies \cite{hewitt_predicting_2024}. 
\\
\\
Below each \textit{free-response} question, participants saw the following:
\begin{quote}
“Please do not include any directly or indirectly personally identifiable information, such as your name or address.”
\end{quote}

\subsection*{Demographics}
\textbf{Source:} Guidance for Researchers When Using Inclusive Demographic Questions for Surveys: Improved and Updated Questions \cite{hughes_guidance_2022}
\begin{itemize}[leftmargin=2cm,itemindent=-1cm]
    \item Demographic Survey: 1, 2, 3, 4, 5, 6, 7, 8, 9, 10, 21, 22, 24
\end{itemize}
\textbf{Source:} 2022 GSS Cross-Section Questionnaire in English \cite{davern_general_2024}
\begin{itemize}[leftmargin=2cm,itemindent=-1cm]
    \item Demographic Survey: 11, 12, 13, 14, 15, 16, 17, 18 19, 20, 23, 25
\end{itemize}

\subsection*{Race and Death Penalty}
\textbf{Source:} Persuasion and Resistance: Race and the Death Penalty in America \cite{peffley_persuasion_2007}
\begin{itemize}[leftmargin=2cm,itemindent=-1cm]
    \item Learning Survey 1: Questions 1, 2
    \item Learning Survey 2: Questions 3, 4
    \item Learning Survey 3: Questions 3, 4
    \item Testing Phase: Questions 6, 7
\end{itemize}

\subsection*{Crime Risks vs. Reward}
\textbf{Source:} Accounting for the Correlation between the Perceived Risks and Rewards to Crime \cite{rebecca_bucci_accounting_2023}
\begin{itemize}[leftmargin=2cm,itemindent=-1cm]
    \item Learning Survey 1: Questions 3, 4, 5, 6
    \item Learning Survey 2: Questions 5, 6, 7, 8
    \item Learning Survey 3: Questions 5, 6, 7, 8
    \item Testing Phase: Questions 8, 9, 10, 11, 12, 13, 14, 15
\end{itemize}

\subsection*{Healthcare Costs}
\textbf{Source:} Public Opinion and Attributions for Health Care Costs \cite{katherine_mccabe_public_2019}
\begin{itemize}[leftmargin=2cm,itemindent=-1cm]
    \item Learning Survey 1: Questions 7, 8
    \item Learning Survey 2: Questions 9, 10, 11
    \item Learning Survey 3: Questions 9
    \item Testing Phase: Questions 16, 17, 18, 19, 20, 21
\end{itemize}

\subsection*{Human Rights Backlash}
\textbf{Source:} Human Rights Shaming, Compliance, and Nationalist Backlash \cite{rochelle_terman_human_2020}
\begin{itemize}[leftmargin=2cm,itemindent=-1cm]
    \item Learning Survey 1: Questions 9, 10, 11, 12
    \item Learning Survey 2: Questions 12, 13, 14
    \item Learning Survey 3: Questions 10, 11
    \item Testing Phase: Questions 22, 23, 24, 25, 26, 27, 28, 29, 30, 31
\end{itemize}

\subsection*{Terrorism Fear Correction}
\textbf{Source:} Can Factual Misperceptions be Corrected? An Experiment on American Public Fears of Terrorism \cite{daniel_silverman_can_2020}
\begin{itemize}[leftmargin=2cm,itemindent=-1cm]
    \item Learning Survey 1: Questions 13, 14, 15, 16
    \item Learning Survey 2: Questions 15, 16, 17, 18
    \item Learning Survey 3: Questions 12, 13, 14, 15
    \item Testing Phase: Questions 32, 33, 34, 35, 36, 37, 38, 39, 40, 41, 42, 43
\end{itemize}

\subsection*{Racial Minority Growth}
\textbf{Source:} Racial Majority \& Minority Group Members' Psychological and Political Reactions to Minority Population Growth \cite{maureen_craig_racial_2017}
\begin{itemize}[leftmargin=2cm,itemindent=-1cm]
    \item Learning Survey 1: Questions 17, 18, 19
    \item Learning Survey 2: Questions 1, 2
    \item Learning Survey 3: Questions 1, 2
    \item Testing Phase: Questions 1, 2, 3, 4, 5
\end{itemize}

\section{LLM Prompt}
\label{prompts}

This prompt was submitted to GPT-4o via OpenAI's API after each learning survey and testing survey submission:
\\

\texttt{
For context, we are currently running a study on how well LLM-powered digital twins are able to answer crowdsourcing surveys based on human input. Our aim is for the digital twin to effectively and accurately mimic the human crowd worker.
\\
You are the digital twin model in this study. I will provide you with the human crowd worker's demographic information and preferences, and your task is to respond to questions as closely as you can to the human crowd worker's preferences in the strict format outlined below. Do not use any outside knowledge other than what I've given you on how the human crowd worker thinks. This task is very important. Do this to the best of your ability. If you do not know the answer to a question based on the human crowd worker's input, output "\$".
\\
All references to the word ``human'' mean the specific ``human crowd worker'' that you are trying to mimic. There is only one human you need to mimic.
\\
For the questions in this survey, you will be given the following information as context:
    1. The demographic info of the human that you are trying to mimic.
    2. The human's answers to the prior survey before this survey.
    3. The LLM-powered digital twin's answers to the same survey that the human answered.
The purpose of this study is for the LLM-powered digital twin to learn from the human's responses to better mimic the human crowd worker. Therefore, the human will answer the same survey that you answer, and you need to try to learn from the human's responses to better understand their opinions and preferences. Whenever the answers from (2) and (3) do not match up, always prioritize the human's answer.
\\
Your task is split into two steps:
\\
Step 1: Write a short paragraph summarizing the key points of the human's demographic information and \\ preferences. IMPORTANT: If you did not receive 
any demographic info, for step 1, simply output ``Did not receive demographic  
information.''
\\
Step 2: Using all of the context information given to you as well as the paragraph you wrote in Step 1, answer the survey questions given to you to the best of your ability following the response format below.
\\
To clarify, the only things you will output back are the paragraph from Step 1 
and the numerical answers from Step 2.  
\\
How to answer survey questions:
    Do your best to predict what the human would choose for each question on a scale of 1 to 7:
    - 1 = Strongly Disagree (or appropriate extreme)
    - 7 = Strongly Agree (or appropriate extreme)
    - Intermediate values (2-6) represent varying degrees between these extremes.
\\
Please format your responses in JSON format as follows:
Sample response structure:
  \{
      "step1": "Your paragraph here",
      "step2": \{
          "1": "1",
          "2": "5",
          "3": "3",
          "4": "2",
          "5": "4",
          "6": "5",
          "7": "1",
          "8": "6",
          "9": "2",
          "10": "1"
      \}
  \}     
\\
}

Following these instructions, a JSON object containing the user's demographic data and all numerical answers from previous learning surveys by both the human and the LLM is submitted.

\end{document}